\documentclass[%
 reprint,%
 amsmath,%
 amssymb,%
 aps,%
 floatfix]{revtex4-2}

\usepackage{graphicx}
\usepackage{mathtools}
\usepackage{dsfont}
\usepackage{bm}
\usepackage[dvipsnames]{xcolor}
\usepackage[colorlinks,%
            pdfusetitle,%
            bookmarksopen=false,%
            urlcolor=Blue,%
            citecolor=Blue,%
            linkcolor=Blue,%
            pdfencoding=auto,%
            psdextra]{hyperref}
\usepackage{physics}
\usepackage{upgreek}
\usepackage{bbm}

\newcommand{\Sr}{\ensuremath{{}^\text{88}\text{Sr}}}
\newcommand{\asciimathunit}[1]{\ensuremath{\,\mathrm{#1}}}

\newcommand{\nm}{\asciimathunit{nm}}
\newcommand{\Hz}{\asciimathunit{Hz}}
\newcommand{\kHz}{\asciimathunit{kHz}}
\newcommand{\MHz}{\asciimathunit{MHz}}
\newcommand{\us}{\ensuremath{\,\upmu \mathrm{s}}}
\newcommand{\ms}{\asciimathunit{ms}}
\newcommand{\s}{\asciimathunit{s}}
\newcommand{\dB}{\asciimathunit{dB}}
\newcommand{\subfiglabel}[1]{\textbf{#1},}
\newcommand{\subfigref}[1]{#1}
\newcommand{\subfigrefbf}[1]{\textbf{#1}}

\def \FourFourxi {\ensuremath{3.8(6)\dB}}
\def \FiveFourteenxi {\ensuremath{3.4(3)\dB}}

\def \FourFourTdark {\ensuremath{26.0\ms}}
\def \FiveFourteenTdark {\ensuremath{54.5\ms}}

\def \FourFourstab {\ensuremath{2.829(4) \times 10^{-15}}}
\def \FiveFourteenstab {\ensuremath{1.087(1) \times 10^{-15}}}

\def \FiveFourteenUltimate {\ensuremath{3 \times 10^{-17}}}

\def \FourFourstabOverSQL {\ensuremath{3.52(1)\dB}}
\def \FourFourstabOverCSS {\ensuremath{3.69(2)\dB}}

\def \FiveFourteenstabOverSQL {\ensuremath{1.94(1)\dB}}
\def \FiveFourteenstabOverCSS {\ensuremath{2.30(1)\dB}}

\def \FourFourCRamseyCSS {\ensuremath{0.97(1)}}
\def \FourFourCRamseySSS {\ensuremath{0.87(1)}}

\def \FiveFourteenCRamseyCSS {\ensuremath{0.96(1)}}
\def \FiveFourteenCRamseySSS {\ensuremath{0.86(1)}}

\begin{document}

\title{Realizing spin squeezing with Rydberg interactions in a programmable optical clock}

\author{William J. Eckner, Nelson \surname{Darkwah Oppong}, Alec Cao, Aaron W. Young, William R. Milner, John M. Robinson, Jun Ye, Adam M. Kaufman}
\affiliation{%
JILA, University of Colorado and National Institute of Standards and Technology,
and Department of Physics, University of Colorado, Boulder, Colorado 80309, USA
}%

\date{\today}

\begin{abstract}

Neutral-atom arrays trapped in optical potentials are a powerful platform for studying quantum physics, combining precise single-particle control and detection with a range of tunable entangling interactions \cite{schleier2010states,gross2017quantum,browaeys2020many}.
For example, these capabilities have been leveraged for state-of-the-art frequency metrology~\cite{bloom2014optical,mcgrew2018atomic, young2020half,bothwell2022resolving} as well as microscopic studies of entangled many-particle states \cite{fukuhara2015spatially, islam2015measuring, kaufman2016quantum, omran2019generation,graham2022multi,bluvstein2022quantum,zhang2022functional}.
In this work, we combine these applications to realize spin squeezing -- a widely studied operation for producing metrologically useful entanglement -- in an optical atomic clock based on a programmable array of interacting optical qubits. 
In this first demonstration of Rydberg-mediated squeezing with a neutral-atom optical clock, we generate states that have almost $4 \, \mathrm{dB}$ of metrological gain.
Additionally, we perform a synchronous frequency comparison between independent squeezed states and observe a fractional frequency stability of \FiveFourteenstab{} at one-second averaging time, which is \FiveFourteenstabOverSQL{} below the standard quantum limit, and reaches a fractional precision at the $10^{-17}$ level during a half-hour measurement.
We further leverage the programmable control afforded by optical tweezer arrays to apply local phase shifts in order to explore spin squeezing in measurements that operate beyond the relative coherence time with the optical local oscillator.
The realization of this spin-squeezing protocol in a programmable atom-array clock opens the door to a wide range of quantum-information inspired techniques for optimal phase estimation and Heisenberg-limited optical atomic clocks~\cite{toth2014quantum, kaubruegger2019variational, kaubruegger2021quantum, kessler2014heisenberg, pezze2020heisenberg}.
\end{abstract}

\maketitle
In the development of quantum enhanced technologies, the field of metrology has emerged as a compelling frontier \cite{pezze2018quantum}.
For example, the use of entangled states of light has already led to enhanced searches for dark matter \cite{backes2021quantum} and improved detection rates in gravitational-wave sensors \cite{tse2019quantum}.
Ground-breaking advances in optical-frequency metrology have also positioned atomic clocks as a promising platform for practical applications of entangled states \cite{ludlow2015optical}, since leading optical clock technologies are now limited by the so-called standard quantum limit (SQL), which is a fundamental bound on the precision of unentangled sensors.
Engineering metrologically useful entanglement could therefore lead to more precise time-keeping, as well as improved studies of fundamental symmetries \cite{sanner2019optical}, searches for dark matter \cite{kennedy2020precision}, and measurements of gravity at smaller length scales \cite{bothwell2022resolving, zheng2022lab}. 

The pursuit of quantum enhancements in optical frequency measurements introduces a variety of experimental hurdles, as the generation and read-out of useful entangled states typically require controlled interactions or collective measurements, high-fidelity atomic-state control, and isolation from noise in order to preserve the fragile quantum correlations that underlie metrological gains~\cite{pezze2018quantum}.
In the face of these challenges, a particularly robust class of entangled states -- known as spin-squeezed states (SSSs) -- has emerged as a powerful and effective resource for achieving sub-SQL performance in neutral-atom clocks operating in the microwave domain~\cite{schleier2010states, cox2016deterministic, greve2022entanglement,braverman2019near,hosten2016measurement, malia2022distributed}.
Pioneering experiments have pushed spin squeezing into the optical domain using collective atom-cavity coupling to generate SSSs on the clock transition in atomic ytterbium~\cite{pedrozo2020entanglement, colombo2022time}.
Recently, a cavity-QED-based strontium lattice clock demonstrated enhanced angular resolution below the SQL, along with a differential clock measurement between two transportable spin-squeezed ensembles with precision below the quantum-projection-noise limit at the~$10^{-17}$ level~\cite{robinson2022direct}. Such experiments with all-to-all cavity-mediated interactions are a promising route toward scalable spin squeezing.
So far, these realizations do not harness microscopic control or detection, or access high-fidelity rotations, which are important ingredients in many new proposals for engineering optimal quantum sensors and can aid reaching performance below the SQL during clock operation~\cite{toth2014quantum, pezze2018quantum, kaubruegger2021quantum, kaubruegger2019variational, kaubruegger2021quantum, kessler2014heisenberg, pezze2020heisenberg}.

\begin{figure*}
\vspace{-1.25em}
\includegraphics[width=\textwidth]{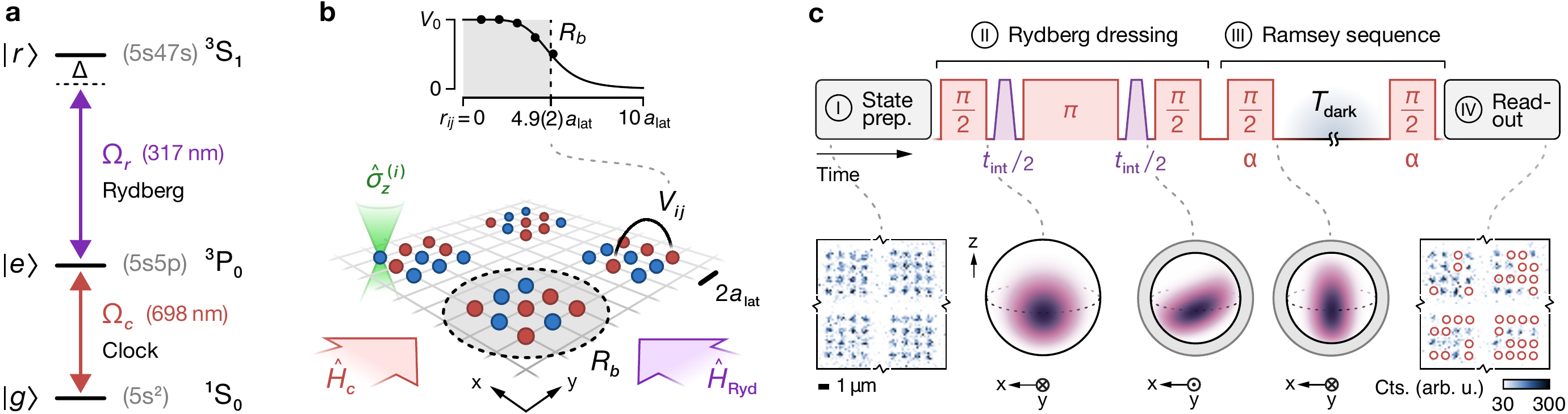}
\vspace{-1.25em}
\caption{\label{fig:1}
    \textbf{Spin squeezing in a Rydberg-dressed array of \Sr{} atoms.}
    \subfiglabel{a}~Atomic states (black lines) and transitions (colored arrows) of \Sr{} relevant for clock interrogation and Rydberg dressing.
    Rabi frequencies $\Omega_{r,c}$ indicate the laser coupling between the different states in the experiment and $\Delta$ denotes the detuning from $\ket{r}$.
    \subfiglabel{b}~Schematic of the experimental setup.
    \Sr{} atoms in the states $\ket{g}$ (blue circles) and $\ket{e}$ (red circles) are trapped in an optical lattice (gray lines), with a lattice spacing of $a_\mathrm{lat}\approx575\nm$, and arranged into multiple subarrays.
    Interactions between atoms at a distance~$r_{ij}$ are described by~$V_{ij}$ shown in the upper plot.
    Here, black circles are data points (error bars are smaller than the marker size) and the solid black line is a numerical fit (see Methods).
    Laser beams coupling the states in panel~\textbf{a} are indicated by colored arrows and correspond to $\hat{H}_c$ and $\hat{H}_\mathrm{\text{Ryd}}$ (see main text).
    Optical tweezers (green double cone, wavelength $515\nm$) enable application of the operator $\hat{\sigma}_z^{(i)}$ to an individual atom~$i$.
    \subfiglabel{c}~Illustration of the experimental sequence for preparing spin squeezed states.
    In the pulse sequences for (II) and (III), clock and Rydberg laser pulses are shown in red and purple, respectively.
    Here, $\alpha$ denotes the relative clock laser phase.
    Note that the pulse duration $t_\mathrm{int}/2$ is visually enlarged by a factor of $\sim10^3$.
    Single-shot images are taken after state preparation (bottom left) and to read out the atomic populations (bottom right).
    Here, the red circles indicate atoms in $\ket{g}$ which are intentionally removed before the final image.
    Note that the distance between the four subarrays is larger than displayed.
    Generalized Bloch spheres (bottom center) illustrate the evolution of the atomic state for a subarray with $N=4\times4=16$ atoms (the reduction of the Bloch-sphere radius is not shown to scale).
}
\end{figure*}

In this work, we experimentally generate and study spin squeezing in a programmable atom-array optical clock to realize measurement performance below the SQL in a differential clock comparison~\cite{seeSupplement}.
The squeezing protocol we use is based on interactions between atoms that are off-resonantly coupled to a Rydberg state with a high principal quantum number~\cite{bouchoule2002spin,gil2014spin}.
Using this technique, known as Rydberg dressing \cite{johnson2010interactions,honer2010collective,jau2016entangling,borish2020transverse, guardado2021quench, zeiher2017coherent}, we are able to observe finite-range interactions in arrays of up to 140 atoms.
Guided by theoretical studies~\cite{gil2014spin, van2021impacts}, we use the resulting Ising-like Hamiltonian to generate spin squeezing on the optical clock transition in $^{88}$Sr (Fig.~\ref{fig:1}\subfigref{a}).
We characterize this squeezing with the Wineland parameter $\xi_W^2$, which serves as an entanglement witness \cite{friis2019entanglement} and quantifies metrological gain \cite{wineland1992spin}.
Assuming an ideal optical local oscillator, one practical way of understanding the Wineland parameter is that a clock with~$N$ atoms and~$\xi_W^2$ will be able to operate with the same precision as an unentangled, SQL-limited (i.e. $\xi_W^2 = 1$) clock with $N / \xi_W^2$ atoms.
Therefore, states with $\xi_W^2 < 1$ ($\xi_W^2  < 0\,\text{dB}$) contain metrologically useful entanglement.
We also perform a first exploration of the impact of squeezing on differential clock measurements at times beyond the atom-laser coherence, for which we leverage local single-qubit gates to impart well-controlled clock phase shifts for unbiased phase estimation~\cite{zheng2022differential,marti2018imaging,young2022tweezer,stockton2007bayesian, seeSupplement}. 

We create spin squeezing between the $^1$S$_0$ ground (denoted $\ket{g}$) and $^3$P$_0$ clock (denoted $\ket{e}$) electronic states (see Fig.~\ref{fig:1}\subfigref{a}), which are of interest due to their long lifetimes, insensitivity to environmental perturbations, and optical carrier frequency.
For example, coherent superpositions of these states can persist on the half-minute timescale~\cite{young2020half,bothwell2022resolving,zheng2022differential}, support long-lived entanglement \cite{schine2022long}, and are the foundation for state-of-the-art neutral-atom optical clocks \cite{bloom2014optical,mcgrew2018atomic}.
We generate spin-squeezed ensembles with $\xi_W^2 = -\FourFourxi$ and $-\FiveFourteenxi$ in subarrays of $N=16$ and $70$ atoms, respectively. We then incorporate this spin squeezing into a differential clock comparison between two independent subarrays of atoms.
For a measurement time of $\tau$, we observe that the fractional-frequency uncertainty in this comparison averages down with a rate of $\FiveFourteenstab/\sqrt{\tau / \rm{s}}$ for $N=70$-atom SSSs, and reaches an ultimate fractional uncertainty below \FiveFourteenUltimate{} after a $27.6 \, $minute-long measurement.
This stability is \FiveFourteenstabOverCSS{} better than the same measurement performed with coherent spin states (CSSs), and \FiveFourteenstabOverSQL{} below the SQL in a differential clock comparison.

The core elements of the programmable clock platform are shown in Fig.~\ref{fig:1}\subfigref{b}, and are based on a recently demonstrated hybrid tweezer-lattice architecture~\cite{schine2022long,young2022tweezer}.
We can trap atoms in both a dynamically configurable optical tweezer array and a collocated two-dimensional (2D) optical lattice, each of which exhibits distinct and enabling features for the work presented here (for more details, see Methods).
In particular, the optical tweezer array allows for rapid initial loading, deterministic rearrangement into nearly arbitrary patterns within the 2D lattice, and the application of controlled, local light shifts, shown schematically in Fig.~\ref{fig:1}\subfigref{b}.
The 2D lattice, on the other hand, offers several thousand sites in which we can perform ground-state cooling, single-site-resolved imaging, and high-fidelity global rotations on the clock transition. 

Our experiments require a combination of global, laser-driven clock rotations -- with a typical Rabi frequency of $\Omega_c \approx 2\pi \times 250 \, \rm{Hz}$ -- and Rydberg-mediated interactions.
We turn on these interactions by applying a high-power $316.9\nm$ laser that addresses the $\ket{e} \leftrightarrow \ket{r} = (5s47s)\,{}^3\text{S}_1$ transition with a typical Rabi frequency $\Omega_r \approx 2\pi \times 5.5 \MHz$ and detuning $\Delta\approx2\pi\times 11\MHz$.
When $\Delta \gg \Omega_r$ (``weak dressing"), this dresses the excited state $\ket{e}$ with an admixture of $\ket{r}$ and creates a new eigenstate $\ket{e_{\rm{dr.}}} \approx \ket{e} - \beta \ket{r}$, where $\beta = \Omega_r / (2\Delta)$.
Because pairs of Rydberg atoms interact through a van der Waals potential with coefficient $C_6$, an effective Hamiltonian for the pseudo-spin states $\{\ket{g}, \ket{e} \}$ can be written as a sum of the two independently controlled terms $\hat{H} = \hat{H}_c + \hat{H}_{\text{Ryd}}$~\cite{henkel2010three, gil2014spin, zeiher2017coherent},
\begin{align}\label{eq:ham}
    \begin{split}
        \hat{H}_c &= \hbar \, \Omega_{c} \hat{S}_x ,\\
        \hat{H}_{\text{Ryd}} &= \frac{1}{4} \sum_{i < j} V_{ij} \hat{\sigma}_z^{(i)} \hat{\sigma}_z^{(j)} + \frac{1}{2} \sum_i \delta_i \hat{\sigma}_z^{(i)},
    \end{split}
\end{align}
where the indices $i$, $j$ label the atoms in the array, and $\hat{\sigma}_{x,y,z}$ are the Pauli operators.
We also denote the collective spin operators $\hat{S}_{\gamma} = \frac{1}{2} \sum_i \hat{\sigma}_{\gamma}^{(i)}$ with \mbox{$\gamma \in \{x,y,z\}$}, and associated Bloch vector $\vec{S} = ( S_x, S_y, S_z)$, where $S_\gamma = \langle \hat{S_\gamma} \rangle$.
The parameter $\delta_i$ describes a longitudinal field term arising in the effective Hamiltonian~\cite{gil2014spin}.
In the weak-dressing limit, the strength of the interactions is given by a soft-core potential $V_{ij} = V_0 / [1 + (r_{ij} / R_b)^6]$ where $r_{ij}$ is the distance between atoms~$i$ and~$j$, $V_0 = \hbar \beta^3 \Omega_r$, and the interaction range is given by $R_b = |C_6 / (2 \Delta)|^{1/6}$.
Using pairs of atoms with variable spacing, we can directly measure the shape of the soft-core potential~\cite{zeiher2017coherent, schine2022long}.
As shown in Fig.~\ref{fig:1}\subfigref{b}, we observe a two-particle interaction that has a spatial dependence consistent with the weak-dressing model, yet with quantitative deviation that is partly attributable to violation of the weak-dressing approximation (see Methods).

The spin-squeezing protocol we use largely follows that proposed in \cite{gil2014spin}, illustrated in Fig.~\ref{fig:1}\subfigref{c}, which applies the interaction Hamiltonian~$\hat{H}_{\text{Ryd}}$ in a spin-echo sequence.
This procedure thereby isolates the interaction $\hat{V}_{\text{int}} = \frac{1}{4} \sum_{i < j} V_{ij} \hat{\sigma}_z^{(i)} \hat{\sigma}_z^{(j)}$, which is applied for a total time~$t_{\text{int}}$ (see Methods).
We can understand the interaction $\hat{V}_{\text{int}}$ by noting that it has a form similar to the one-axis twisting (OAT) Hamiltonian $\hat{H}_{\text{OAT}} = \chi \hat{S}_z^2 \,$, which has been widely studied as a generator of spin squeezing \cite{kitagawa1993squeezed, leroux2010implementation}.
In the limit where $R_b$ is much larger than the maximum inter-atomic distance, $\hat{V}_{\text{int}} \approx \hat{H}_{\text{OAT}}$. OAT acts by redistributing quantum noise between two non-commuting observables which can result in a metrological gain when measuring the observable with lower noise; we can see that $\hat{V}_{\text{int}}$ similarly ``squeezes'' the quantum noise in the system (Fig.~\ref{fig:1}\subfigref{c}), even in the limit where $R_b$ is smaller than the maximum inter-atomic distance~\cite{gil2014spin}. 

We study the squeezing generated with $\hat{V}_{\text{int}}$ for small systems by preparing atoms in four square subarrays of $N=4 \times 4 = 16$ atoms (with an inter-atomic spacing of $2 \, a_{\text{lat}}$).
By ensuring that the separation between subarrays is larger than the interaction range $R_b$, we can treat them as independent populations.
This allows us to extract information about their intrinsic quantum noise by performing differential comparisons, which reject most forms of technical noise~\cite{schulte2020prospects}.
The observable for this differential measurement is $\hat{d}^{(AB)}_z = \hat{S}^{(A)}_z / N_A - \hat{S}^{(B)}_z / N_B$, where $A$ and $B$ each label a subarray of atoms, with atom numbers $N_A$ and $N_B$, respectively (see Methods).
We can assume that each subarray is a preparation of the same atomic state, and then treat all measurements of~$\hat{d}^{(AB)}_z$ as a probe of the same observable, which we call~$\hat{d}_z$.
The variance in~$\hat{d}_z$, denoted~$\sigma_{\alpha}^2$, generally depends on the measurement quadrature, which is set by the angle $\alpha$, and describes the orientation of the atomic noise distribution.
However, in the absence of entanglement or technical noise, $\sigma_{\alpha}^2$ will be independent of $\alpha$, and governed by quantum projection noise (QPN), given by
\begin{align}
    \sigma_{\text{QPN}}^2 = \text{Var}\Biggl{[}\frac{\hat{S}_z^{(A)}}{N_A} - \frac{\hat{S}_z^{(B)}}{N_B}\Biggr{]}_{\text{CSS}} = \frac{1}{2N},
\end{align}
where we assume $N_A, N_B = N$, as we prepare $N_A \approx N_B$ for all experiments. $\text{Var}[.]_{\text{CSS}}$ denotes the variance of an ideal CSS with $S_z = 0$ for both subarrays $A$ and $B$. 

\begin{figure}
\centering
\includegraphics[width=\linewidth]{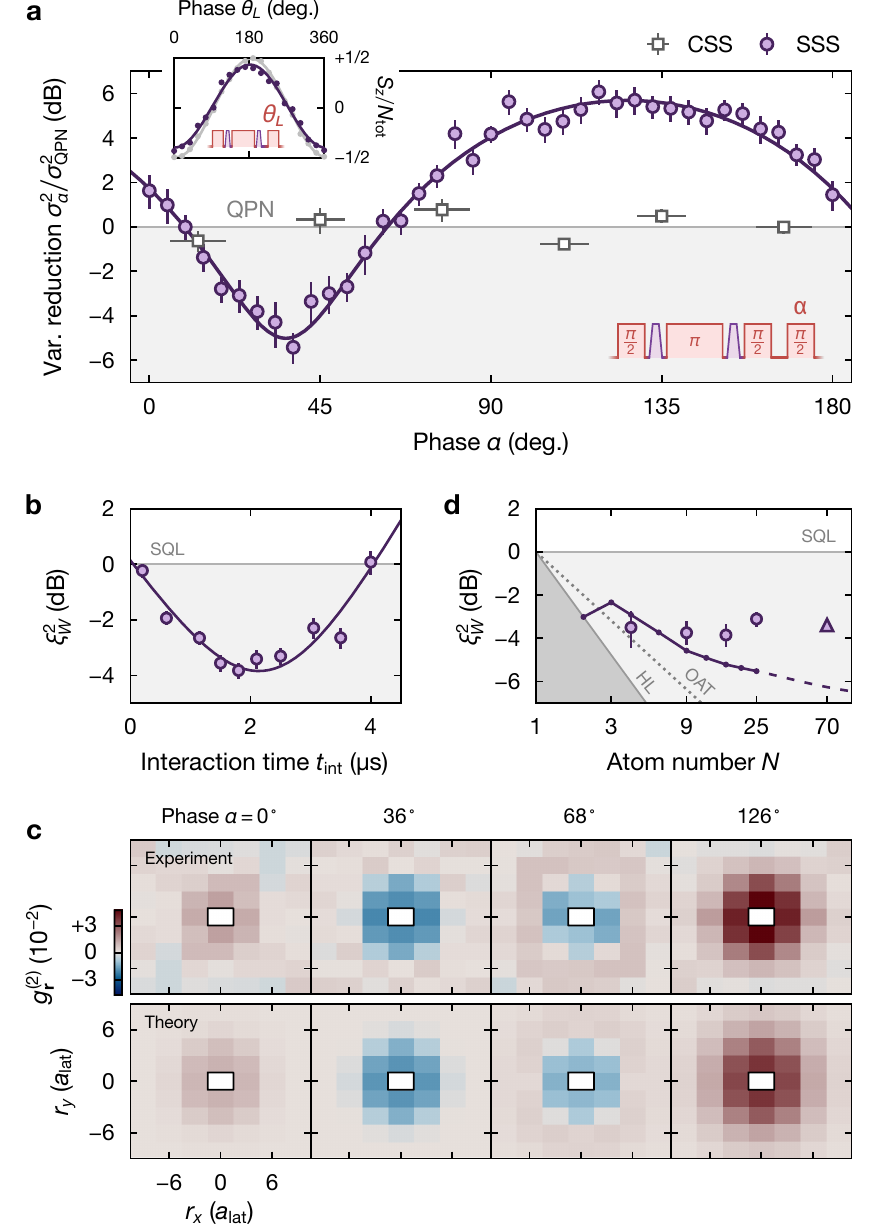}
\caption{
    \label{fig:2}
    \textbf{Characterization of spin squeezing and finite-range interactions.}
    Unless noted otherwise, the displayed data correspond to $N=4\times4=16$ atoms.
    \subfiglabel{a}~Variance reduction $\sigma_\alpha^2 / \sigma_\mathrm{QPN}^2$ of the coherent spin state (CSS, gray squares) and the spin-squeezed state (SSS, purple circles) for variable phase $\alpha$ of the final clock-laser pulse and $t_\mathrm{int}=2.4\us$.
    Data points for the CSS are binned and the solid purple line is a guide to the eye.
    The top left inset shows $S_z / N$ of the CSS (gray points) and SSS (purple points) for variable phase~$\theta_L$, and averaged over subarrays.
    The solid lines represent fits yielding the contrast $C=0.97(1)$ [$0.83(2)$] for the CSS [SSS].
    \subfiglabel{b}~Wineland squeezing parameter $\xi^2_W$ (purple circles) measured for variable $t_\mathrm{int}$.
    The dark purple line shows an empirical fit to determine the optimal $\xi_W^2$.
    \subfiglabel{c}~Two-particle correlator $g^{(2)}_\mathbf{r}$ (see main text) in a subarray with $N=5\times14=70$ atoms and close to the optimal interaction time, $t_\mathrm{int} = 1.6\us$, for variable phase $\alpha$ and horizontal (vertical) displacement $r_x$ ($r_y$).
    The bottom rows show the weak-dressing theory at optimal interaction time for an offset angle $\alpha \rightarrow \alpha - 9.9^\circ$ determined from a fit to the experimental data.
    We note that the optimal interaction time differs for the experiment and theory shown here (see Methods).
    \subfiglabel{d}~Optimal Wineland squeezing parameter $\xi_W^2$ (large purple markers) for variable atom number $N = \sqrt{N} \times \sqrt{N}$ shown on a logarithmic axis.
    The triangle marker corresponds to a rectangular array with $N=5\times14$.
    Small purple circles and lines show the theoretical prediction based on weak-dressing (dashed line indicates subarrays with $N=5\times m$ for $N>25$).
    The solid and dotted gray lines indicate the Heisenberg limit (HL) and the asymptotic scaling for the one-axis twisting Hamiltonian (OAT), respectively.
    }
\end{figure}

As shown in Fig.~\ref{fig:2}\subfigref{a}, we can measure $\sigma_\alpha^2$ after applying the squeezing protocol to $N = 4\times 4$ subarrays. 
We find that the ratio $\sigma_{\alpha}^2 / \sigma_{\text{QPN}}^2$ oscillates sinusoidally with~$\alpha$, and dips below unity near $\alpha \approx 30^\circ$.
This measurement demonstrates that SSSs in the system have noise below the QPN limit.
However, we must also ensure that this variance reduction is not offset by a reduction in contrast, which could in net reduce the signal to noise ratio~\cite{seeSupplement}.
To verify this, we measure the contrast~$C$ of the Ramsey fringe associated with each of these states, and thus the magnitude of the single-ensemble Bloch vector~$S = CN/2$.

From the measured quantities $\sigma_{\alpha}^2$ and $C$, we can determine the Wineland squeezing parameter
\begin{equation}
    \xi_W^2 = \frac{N}{2 S^2} \text{Var}{\left[\hat{S}_z^{(A)} - \hat{S}_z^{(B)}\right]}_{ \text{min}} =  \frac{1}{C^2} \frac{\sigma_{\text{min}}^2}{\sigma_{\text{QPN}}^2} \, ,
\end{equation} 
where $\sigma_{\text{min}}^2$ denotes the minimum of $\sigma_{\alpha}^2$ and $\text{Var}[\cdot]_{ \text{min}}$ is the minimum variance with respect to $\alpha$. We determine the optimal~$\alpha$ by fitting a cosine to the signal~$\sigma_{\alpha}^2$ (see Fig.~\ref{fig:2}\subfigref{a}).
The parameter $\xi_W^2$ is then calculated from the variance of an additional, high-statistics dataset taken at the optimal $\alpha$, as well as the fitted contrast $C$ (see inset of Fig.~\ref{fig:2}\subfigref{a}).
Figure~\ref{fig:2}\subfigref{b} shows $\xi_W^2$ versus interaction time~$t_{\text{int}}$ for $N=4 \times 4$ subarrays.
By fitting the measured Wineland parameter $\xi_W^2$ versus $t_\mathrm{int}$, we observe a minimum value of $\xi_W^2 =-3.8(6)\,\mathrm{dB}$, which is comparable to state-of-the-art demonstrations in other optical clocks~\cite{robinson2022direct,colombo2022time}.

\begin{figure*}
    \includegraphics[width=\linewidth]{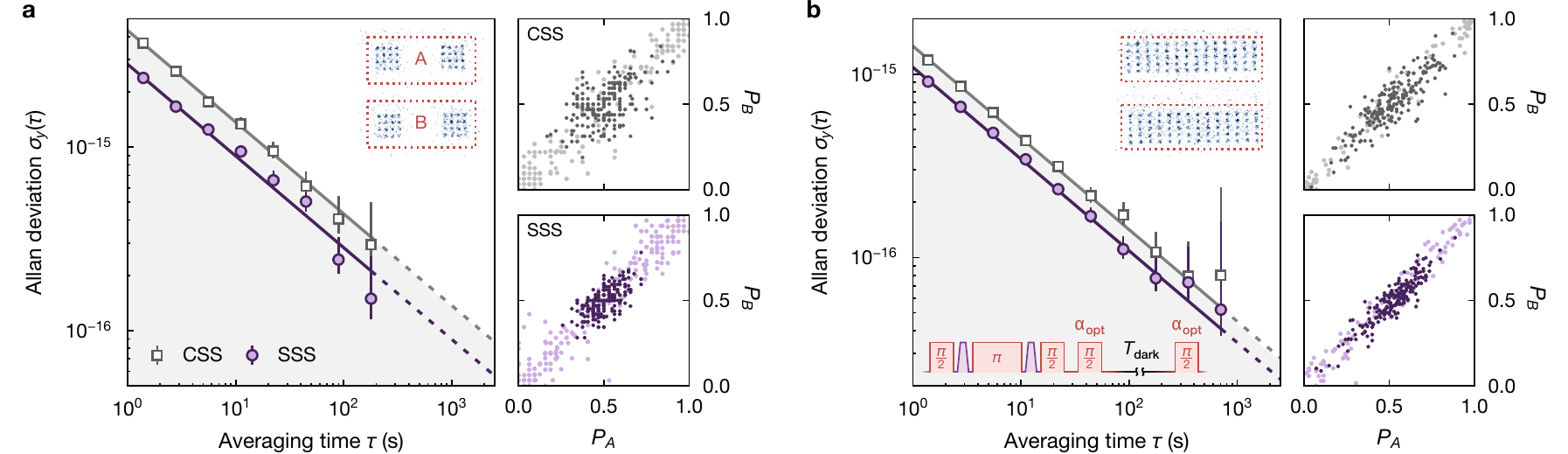}
    \caption{
    \label{fig:3} 
    \textbf{Atom-atom stability for coherent spin states and spin-squeezed states.} 
    Overlapping Allan deviation for the differential clock comparison between two subarrays after preparing a coherent spin state (CSS, gray squares) or a spin-squeezed state (SSS, purple circles).
    The subarrays $A$ and $B$ are illustrated in the top right inset of each main panel with \subfiglabel{a}~$N_{\{A,B\}}=2 \times (4\times4) = 32$ and \subfiglabel{b}~$N_{\{A,B\}}=5\times 14 = 70$.
    The atoms are interrogated with a dark time of \subfiglabel{a}~$T_\mathrm{dark} =\FourFourTdark{}$ and \subfiglabel{b}~\FiveFourteenTdark{} between two $\pi/2$ pulses with the laser phase $\alpha_\mathrm{opt}$ optimized for the measurement quadrature with lowest noise (see bottom left inset of main panel in~\subfigref{b}).
    Solid lines show numerical fits to the data yielding differential stabilities of \mbox{$\FourFourstab{}/\sqrt{\tau / \rm{s}}$} and \mbox{$\FiveFourteenstab{}/\sqrt{\tau / \rm{s}}$} for the SSS in \subfigrefbf{a} and \subfigrefbf{b}, respectively.
    These stabilities are respectively \FourFourstabOverSQL{} and \FiveFourteenstabOverSQL{} below the SQL for a differential clock comparison~\cite{seeSupplement}, and correspond to a \FourFourstabOverCSS{} and \FiveFourteenstabOverCSS{} enhancement over the CSS.
    The two right panels in \subfigrefbf{a} and \subfigrefbf{b} show the excitation probabilities $P_A$ and $P_B$ of subarrays $A$ and $B$, respectively.
    The light purple and light gray points correspond to a separate measurement, in which the phase of the final $\pi/2$-pulse is varied over $360^\circ$ to determine the contrast $C=\FourFourCRamseyCSS{}$ [\FourFourCRamseySSS{}] in \subfigrefbf{a} and $C=\FiveFourteenCRamseyCSS{}$
    [\FiveFourteenCRamseySSS{}] in \subfigrefbf{b} for the CSS [SSS].
    The dark purple and gray points correspond to the first~$200$ points of the raw data from which the corresponding Allan deviations in the main panels are calculated.
    The spread of the data points illustrates how much of the Ramsey fringe is sampled when locking the clock laser to the atomic signal (see Methods).
    Note how the reduced variance of the SSS compared to the CSS can be seen close to $P_A \approx P_B \approx 0.5$ along the anti-diagonal of this plot.}
\end{figure*}

Using site-resolved imaging, we can also probe the microscopic structure of the generated states by analyzing the two-particle correlator $g^{(2)}_{\textbf{r}} = \frac{1}{4}\langle \hat{\sigma}_z^{(i)} \hat{\sigma}_z^{(j)} \rangle - \frac{1}{4}\langle \hat{\sigma}_z^{(i)} \rangle \langle \hat{\sigma}_z^{(j)} \rangle$, where $\textbf{r}$ is a spatial displacement vector between lattice sites~$i$ and~$j$.
We measure $g^{(2)}_{\textbf{r}}$ in larger subarrays to reduce finite-size effects.
For $N = 5 \times 14=70$ (with spacing~$3\, a_{\text{lat}}$ along $x$ and $2\, a_{\text{lat}}$ along $y$), we find that the measured $g^{(2)}_{\textbf{r}}$ agrees qualitatively with theoretical predictions at the optimal interaction time (see Methods).
In particular, we observe correlations that extend over a range that is similar to the characteristic length scale $R_b\approx5a_\mathrm{lat}$ of the interaction potential $V_{ij}$ shown in Fig.~\ref{fig:1}\subfigref{b}.
For $|\textbf{r}| < R_b$, we observe correlations that change from negative to positive as a function of~$\alpha$.
As expected, the $\alpha$ with minimum Wineland parameter~$\xi_W^2$ exhibits strong negative correlations.

An important question concerns how the Wineland parameter~$\xi_W^2$ changes with increasing atom number~$N$.
For the finite-range interactions realized by Rydberg dressing, we expect~$\xi_W^2$ to saturate when the mean interatomic distance becomes much larger than~$R_b$~\cite{gil2014spin}.
To probe this regime, we perform additional measurements of the optimal~$\xi^2_W$ with $N=4$, $9$, $25$ and $70$-atom subarrays.
For each, we employ an empirical fit (see Fig.~\ref{fig:2}\subfigref{b}) to determine the optimal Wineland parameter~$\xi_W^2$, shown in Fig.~\ref{fig:2}\subfigref{d} (see Methods).
We do not observe a strong dependence on $N$ in the achievable squeezing~$1/\xi_W^2$, which saturates to~$\approx4\,\mathrm{dB}$ for~$N = 9, 16$, and is slightly reduced for larger subarrays of $N = 25$ and $70$. 
By contrast, the theoretical prediction for~$\xi^2_W$ given $\hat{H}_\mathrm{\text{Ryd}}$ (see purple line in Fig.~\ref{fig:2}\subfigref{d}) continuously improves over the experimentally relevant~$N$ and deviates by more than~$2\,\mathrm{dB}$ from the experimental result at~$N=70$. 
These deviations could originate from unitary dynamics in the system that are not captured by Eq.~\eqref{eq:ham}.
Here, it is important to note that we operate with a relatively large~$\beta \approx 0.25$, which we empirically find to maximize the achievable squeezing.
In this regime, the two-body interaction strengths $V_{ij}$ deviate from the predictions of weak dressing (see Methods), and collective interactions can play an elevated role~\cite{honer2010collective,henkel2010three}.
For example, we find a significant deviation in the dynamics between experiment and theory for $N=9$ when using Eq.~\eqref{eq:ham}, but not when considering exact diagonalization of the full three-level Rydberg Hamiltonian (see Methods).
In addition to altered unitary dynamics, observations in larger subarrays might be impacted by non-unitary dynamics in the form of collective dissipation~\cite{zeiher2016many, boulier2017spontaneous, guardado2021quench, young2018dissipation, festa2022blackbody, seeSupplement}.

So far, we have focused on the preparation of SSSs. 
Next, we benchmark their performance in a synchronous optical-frequency comparison between independent atomic ensembles, labeled $A$ and $B$ (see insets in Fig.~\ref{fig:3}).
We interrogate both CSSs and SSSs in a Ramsey-interferometry sequence with a variable dark time $T_{\text{dark}}$ (see diagram in Fig.~\ref{fig:3}\subfigref{b}). 
During the dark time, the phases of ensembles $A$ and $B$ precess at their angular clock frequencies $\omega_{A}$ and $\omega_{B}$, respectively. 
At the end of the dark time, we then measure the differential phase $\phi$ between the two ensembles, which is related to the previously defined observable $\hat{d}_z^{(AB)}$ by 
\begin{align}
    d^{(AB)}_z \approx \frac{C}{2} \phi = \frac{C}{2} \left( \omega_{A} - \omega_{B} \right) T_{\text{dark}},
\end{align}
when $\phi\ll1$ (see Methods). For ensembles with $N = 32$ (each comprised of two $4 \times 4$ subarrays) and $T_{\text{dark}} = \, $\FourFourTdark{} we observe a fractional frequency stability of $\FourFourstab/\sqrt{\tau / \rm{s}}$ between two SSSs (Fig.~\ref{fig:3}\subfigref{a}).
This corresponds to a \FourFourstabOverSQL{} enhancement over the SQL in a differential clock comparison~\cite{seeSupplement}, and a \FourFourstabOverCSS{} improvement compared to the same measurement performed with CSSs.
With ensembles of size $N = 5 \times 14 = 70$ and $T_{\text{dark}} = \FiveFourteenTdark{}$ we realize a stability of $\FiveFourteenstab{}/ \sqrt{\tau / \rm{s}}$ between two SSSs. Assuming the data continue to average as white frequency noise, this implies a final instability below \FiveFourteenUltimate{} when extrapolated to the full measurement time of 27.6 minutes.
This stability is \FiveFourteenstabOverSQL{} [\FiveFourteenstabOverCSS{}] below the SQL [CSS] for a differential clock comparison~\cite{seeSupplement}.
The smaller metrological gain for $N=70$ could be attributed to a slightly reduced $1/\xi_W^2$ as observed in Fig.~\ref{fig:2}\subfigref{d}.
However, the larger-atom-number arrays still allow us to reach a lower absolute measurement uncertainty at fixed averaging time.
To the best of our knowledge, these measurements are the first to achieve a fractional-frequency precision below the SQL for a differential clock comparison in a neutral-atom optical clock~\cite{seeSupplement}. 

One can extend differential frequency comparisons beyond the atom-laser coherence time~\cite{marti2018imaging, young2020half, zheng2022differential, bothwell2022resolving}.
In this regime, the phase of the second $\pi/2$ pulse in a Ramsey interferometer is completely randomized.
However, the measured clock-state fractions $P_A$ and $P_B$ of two ensembles $A$ and $B$ can be plotted parametrically, and trace out an ellipse with an opening angle set by the differential phase~$\phi$.
Given an appropriate atomic noise model, maximum-likelihood estimation (MLE) can be employed to directly measure~$\phi$.
We refer to this as ``ellipse fitting''~\cite{stockton2007bayesian}. 

\begin{figure*}
\includegraphics[width=0.67\linewidth]{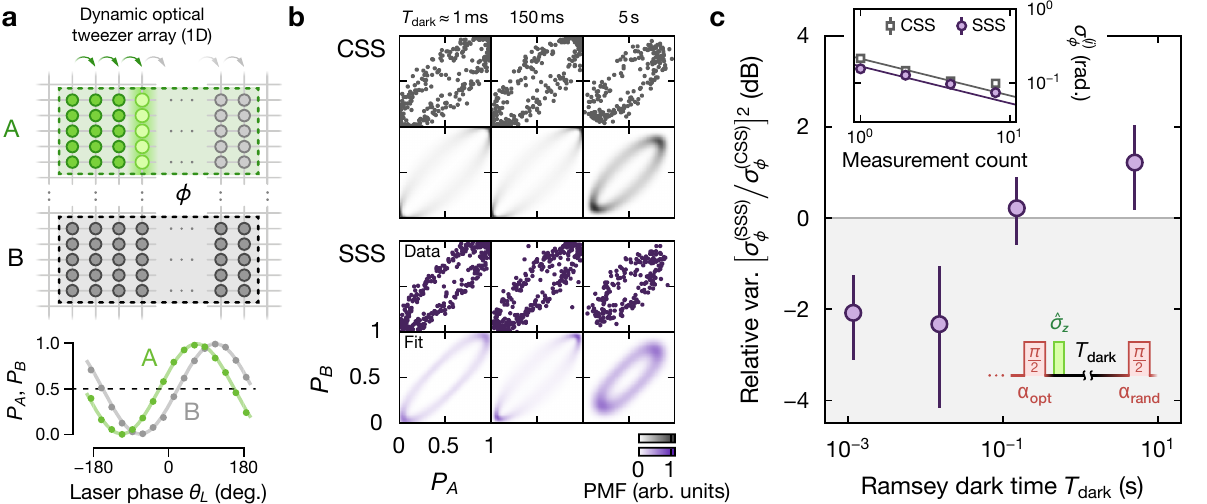}
\caption{
\label{fig:4} 
\textbf{Exploring metrology with spin-squeezed states in the limit of randomized atom-laser phase.}
In all panels, we study two subarrays with $N=5\times 14=70$ each.
\subfiglabel{a}~Schematic of the phase shifting procedure.
Dynamic optical tweezers (light green circles) collocated with the optical lattice (gray lines) apply local operators $\hat{\sigma}_z^{(i)}$ onto the atoms (green and gray circles).
At the bottom, together with a fit (solid lines), an example Ramsey measurement of the phase shift~$\phi$ between subarrays A and B is shown.
\subfiglabel{b}~Measurements of the probabilities $P_A$ and $P_B$ for variable Ramsey dark time $T_\mathrm{dark}=1.16\ms$, $150\ms$, and $5\s$.
Here, the relative phase shift takes a value of~$\phi\approx30^\circ$ ($T_\mathrm{dark}<5\s$)  or~$\phi\approx40^\circ$ ($T_\mathrm{dark}=5\s$).
The first and third rows show the experimental data for the coherent spin state (CSS, gray squares) and the spin-squeezed state (SSS, purple points), respectively. The other rows show the numerically fitted probability mass functions (PMFs, see Methods).
\subfiglabel{c}~Purple circles correspond to the inferred relative variance of the SSS compared to the CSS for variable dark time $T_\mathrm{dark}$.
Each data point is determined by comparing the Allan deviation $\sigma_\phi$ of the estimated-phase uncertainty after averaging over a variable number of measurements.
Here, the inset shows an example for $T_\mathrm{dark}=1.16\ms$ (error bars smaller than the marker size), and the bottom right schematic illustrates part of the pulse sequence in this measurement.}
\end{figure*}

Next, we explore the ellipse-fitting approach using two $N=70$ SSSs. We repeat this measurement for a few different Ramsey dark times and compare the precision to that achieved with comparable CSSs.
One consideration in ellipse fitting is that measurements of $\phi \approx 0$ have biased results~\cite{marti2018imaging, seeSupplement}. 
In order to operate away from this point, we use the local control afforded by optical tweezer arrays to apply a homogeneous phase offset of $\phi\approx30^\circ$ to one of the two ensembles (see Fig.~\ref{fig:4}\subfigref{a}, Methods).
To address residual drift in this phase-shifting protocol (see Methods), we interleave measurements with the CSS and SSS so that they probe the same $\phi$.
In addition, we intentionally randomize the laser phase of the final $\pi/2$ pulse in the interferometer, which ensures uniform sampling of the ellipse traced out by $P_A$ and $P_B$, independent of the atom-laser coherence time.
Fig.~\ref{fig:4}\subfigref{b} shows how the data cluster near an ellipse with an opening angle of~$\approx30^\circ$.
The rows below the data in Fig.~\ref{fig:4}\subfigref{b} depict the fitted noise distributions~\cite{seeSupplement}, which show good agreement with the experimental data. 

To extract the measurement uncertainty achieved with CSSs and SSSs, we calculate an Allan deviation~$\sigma_\phi$ of the measured phase~$\phi$ through a jackknifing procedure~\cite{marti2018imaging, seeSupplement}.
For this, we choose to parameterize the Allan deviation in terms of number of measurements (as opposed to averaging time $\tau$) due to the interleaved operation of ellipse measurements (see Methods).
The inset in Fig.~\ref{fig:4}\subfigref{c} displays one such Allan deviation obtained for $T_{\rm{dark}} = 1.16 \, \ms$. 
At this short dark time, the SSS provides an inferred measurement uncertainty that is about~$2\dB$ better than that reached with a CSS. 
These results suggest that SSSs could improve the precision of ellipse-fitting measurements, yet there remain open theoretical questions regarding the choice of statistical model for the SSSs (see Methods). 
Furthermore, indications of a potential enhancement are gone by $T_{\text{dark}} = 150 \, \text{ms}$. 
Therefore, this procedure and the associated modeling do not currently allow for higher precision differential Ramsey measurements than those presented in Ref.~\cite{young2020half}.
We leave a detailed study of the technical limitations on squeezing lifetime and how this performance could be extended to longer dark times as a subject for future work.
However, we speculate that inhomogeneous light shifts from the optical lattice and the consequent dephasing could play an important role~\cite{schine2022long}.

In summary, we have employed a Rydberg-dressing protocol to generate up to \FourFourxi{} of spin squeezing on the optical clock transition in \Sr{}.
This has allowed us to perform a synchronous clock comparison with a stability that is up to \FourFourstabOverSQL{} below the SQL.
The demonstrated protocol establishes an effective new approach for reaching entanglement-enhanced optical atomic clocks, and is compatible with other existing experiments~\cite{campbell2017fermi, mcgrew2018atomic,madjarov2019atomic,zheng2022differential}.
Looking to the future, a number of questions and avenues of investigation remain.
Although the CSS and SSS maintain relative atomic coherence for long Ramsey times out to 5 seconds, as noted, we observe signs of an inferred enhancement only at short dark times; correcting this disparity could yield improvements, particularly with refined modeling of the underlying distribution of the experimentally produced entangled states.
Another important consideration is whether the reported stability enhancements can be combined with state-of-the-art accuracy. 
To this end, the switchability of the Rydberg interactions allows the entangling operations and Ramsey-based metrology to be fully decoupled, which should reduce systematic effects related to Rydberg excitations.
This work sets the stage for fundamental investigations of more sophisticated protocols for generating metrologically useful entangled states for quantum sensing, including protocols that leverage dynamics under more complex spin models~\cite{young2022enhancing,block2023universal}, Floquet engineering \cite{geier2021floquet}, or variationally optimized quantum circuits, even in the regime where the resulting many-body dynamics are challenging to simulate with classical resources~\cite{kaubruegger2019variational, kaubruegger2021quantum, marciniak2022optimal}.
Lastly, the single-particle readout and rearrangement demonstrated here could be used to perform mid-circuit measurements in an entangled optical atomic clock to reach Heisenberg-limited performance that is robust to local oscillator noise~\cite{bowden2020improving, kessler2014heisenberg, pezze2020heisenberg, bluvstein2022quantum}

\textit{Note}: During completion of this work, we became aware of related works using Rydberg interactions in a tweezer-array platform \cite{bornet2023scalable} and long-range interaction in an ion string \cite{franke2023quantumenhanced}.

\begin{acknowledgments}
We acknowledge earlier contributions to the experiment from M.~A. Norcia and N. Schine as well as fruitful discussions with S.~Geller, R.~B. Hutson, W.~F. McGrew, S.~R. Muleady, A.~M. Rey, N. Schine, M. Schleier-Smith, J.~K. Thompson, J.~T. Young, and P. Zoller.
The authors also wish to thank S. Geller, S.~R. Muleady, J.~K. Thompson, and P. Zoller for careful readings of the manuscript and helpful comments.
In addition, we thankfully acknowledge helpful technical discussions and contributions to the clock laser system from A. Aeppli, D. Kedar, K. Kim, B. Lewis, M. Miklos, Y.~M. Tso, W. Warfield, L. Yan, Z. Yao.
This material is based upon work supported by the Army Research Office (W911NF-19-1-0149, W911NF-19-1-0223), Air Force Office for Scientific Research (FA9550-19-1-0275), National Science Foundation QLCI (OMA-2016244), U.S. Department of Energy, Office of Science, National Quantum Information Science Research Centers, Quantum Systems Accelerator, and the National Institute of Standards and Technology.
We also acknowledge funding from Lockheed Martin.
W.J.E. acknowledges support from the NDSEG Fellowship; N.D.O. acknowledges support from the Alexander von Humboldt Foundation; and A.C. acknowledges support from the NSF Graduate Research Fellowship Program (Grant No. DGE2040434).
\end{acknowledgments}

\vspace{2em}
\section*{Methods}

\subsection{Array initialization}
In this work, atom arrays in the optical lattice are initialized via a combination of stochastic loading, detection, deterministic rearrangement with optical tweezers, and high fidelity optical cooling.
First, an array of $515\nm$ tweezers is stochastically loaded from a cold atomic cloud.
Light assisted collisions result in an occupation of $0$ or $1$ atoms in each tweezer with approximately equal probability~\cite{norcia2018microscopic}.
These atoms are implanted into a single 2D layer of a three-dimensional (3D) optical lattice operated close to the clock-magic wavelength of $813.4\nm$, and imaged with a combined loss and infidelity of $<0.5\%$~\cite{schine2022long, young2022tweezer}. 
Based on these images, and using the optical tweezers, the atoms are rearranged into nearly arbitrary patterns in the lattice~\cite{young2023atomic2, kumar2018sorting, endres2016atom, barredo2016atom}.
The per-atom success probability for filling a given target pattern can be as high as $99.5\%$; however, $98\%$ is typical for the data appearing throughout this work.
After rearrangement, an additional image confirms that the target pattern has been prepared successfully.
Note that we do not always enforce that the atom array is free of defects, as summarized in the section on \textit{Post-selection}.
Finally, the rearranged atoms are cooled to their 3D motional ground state via resolved sideband cooling on the $\rm ^1S_0 \leftrightarrow {^3P}_1$ transition~\cite{schine2022long, young2022tweezer, young2023atomic2}.

In all measurements presented in Figs.~2 to 4, we prepare the atomic array with a square or rectangular pattern that corresponds to two or four individual subarrays of size $N=4$ up to $70$ atoms.
By spacing the subarrays sufficiently far apart ($\geq 12a_\mathrm{lat}$), we can treat them as independent atomic ensembles.
The spacing between atoms in a subarray is generally chosen to be $2a_\mathrm{lat}$.
For $N=4$ and $70$, the spacing is increased to $(3, 2) a_\mathrm{lat}$ along the $(x, y)$ directions.
We note that this choice is motivated by a slightly improved fidelity of the array initialization, but does not significantly affect the attainable squeezing performance.

\subsection{Post-selection}
For the data in Figs.~\ref{fig:1} and~\ref{fig:2}, we post-select on the initial filling of the atom array after initialization (see \textit{Array initialization}).
The post-selection criterion for all data sets is an initial filling of $\geq 92\%$, which corresponds to full filling for the probed subarray sizes with $N\leq9$.
Note that the criterion is not applied globally, but on the level of individual subarrays for the data shown in Fig.~\ref{fig:2}.
We do not post-select on initial fill for the stability and ellipse-fitting data sets shown in Fig.~\ref{fig:3} and Fig.~\ref{fig:4}.
However, a single shot with zero initial fill is removed from the data set for the CSS shown in Fig.~\ref{fig:3}b.
For Fig.~\ref{fig:4}\subfigref{b},\subfigref{c} -- and in ascending order with Ramsey dark time -- the SSS [CSS] data have (zero, one, two, one) [(zero, two, two, one)] shots with zero initial fill; these are removed from the corresponding data sets.
Since we calculate the Allan deviation versus number of binned data points~$M$ (as opposed to averaging time $\tau$), neglecting these points in the analysis does not affect the inferred stabilities.

\subsection{Clock rotations and state detection}
After initializing the atom array,
a magnetic field of~$\approx 275 \, \rm{G}$  is turned on to allow the $\ket{g} \leftrightarrow \ket{e}$ transition to be resonantly driven with a typical  Rabi frequency of  $\Omega_c \approx 2\pi \times 250\Hz$ \cite{taichenachev2006magnetic, schine2022long}.
We note that for the data shown in Fig.~\ref{fig:1}b and the $N=2\times2=4$ data in Fig.~\ref{fig:2}d, we employ a smaller magnetic field of~$\approx 55\,\mathrm{G}$.
The ultra-narrow clock laser is stabilized to a cryogenic silicon cavity, as described in \cite{matei20171, oelker2019demonstration}. 
Arbitrary clock rotations can be performed by controlling the duration and phase of pulses from this drive laser using an acousto-optic modulator.
We typically measure a $\pi$-pulse fidelity of $\geq 99\%$.
After preparing a SSS or CSS and interrogating the atoms, we detect their electronic state.
To this end, we apply $461\nm$ blowaway light, resonant with the dipole-allowed $\ket{g} \leftrightarrow \mathrm{{^1}P_1}$ transition, which heats $\ket{g}$ atoms out of the trap.
This procedure projects each atom into either the $\ket{g}$ state (detected as loss) or $\ket{e}$ state (detected as survival).
For further details on state detection and imaging, see Ref.~\cite{schine2022long}. 

In order to sample the squeezed quadrature of a SSS, we need to align it with projective measurements of the $\hat{S}_z$-basis states $\ket{g}, \ket{e}$.
To achieve this, we change the phase of the drive $\Omega_c$ by a variable angle $\alpha$ after the Rydberg-dressing pulse sequence (see II in Fig.~\ref{fig:1}{c}).
At this stage the Bloch vector is aligned parallel to the $z$-axis, i.e., $\vec{S} = \pm S(0,0,1)$, and a phase change of the drive can be understood as a global $\hat{S}_z$-rotation by $\alpha$, which does not change the magnitude or direction of the Bloch vector.
After applying another $\pi/2$ pulse, the atomic noise distribution is rotated by $\alpha$ with respect to the equatorial plane of the generalized Bloch sphere (as illustrated in Fig.~\ref{fig:1}\subfigref{c}). 
To maximize the metrological gain in stability and ellipse-fitting measurements, we experimentally determine the optimal $\alpha$ before each measurement and choose the clock laser phase for the final two $\pi/2$ pulses in the Ramsey sequence appropriately (see III in Fig.~\ref{fig:1}c).

\subsection{Local $\hat{\sigma}_z$ operations}
We use $515\nm$ optical tweezers to introduce locally controlled light shifts across the array. These operations can also be understood as local $\hat{\sigma}_z$ rotations.
Note that in combination with arbitrary global single qubit rotations, this technique provides access to a universal set of single qubit gates.
For data presented in Fig.~\ref{fig:4}, and as described in the main text, we demonstrate this control by creating a homogeneous light shift across a 70-atom subarray. To realize this operation, a single column of tweezers is turned on and the desired light shift is applied to one column of atoms. This is iterated column-by-column across the 2D array (see Fig.~\ref{fig:4}a).
The primary motivation for only applying 1D columns of tweezers at any given time is to ensure that we have a sufficient number of degrees of freedom to independently and arbitrarily tune the phase shift at each site. For a further discussion of the performance of this protocol, see Ref.~\cite{seeSupplement}.

\subsection{Rydberg drive and parameters}
Our ultraviolet laser system for addressing the \mbox{$\ket{e} \leftrightarrow \ket{r}$} transition is detailed in Ref.~\cite{schine2022long}.
We switch on (off) this laser by simultaneously ramping $\Omega_r$ to its maximum (minimum) value and $\Delta$ to its minimum (maximum).
Typically, $\Omega_r$ ramps from $0$ to $\approx 5.5 \, \text{MHz}$, and $\Delta$ ramps from $\approx 2 \pi \times 33 \,\text{MHz}$ to $\approx 2 \pi \times 11 \, \text{MHz}$.
These ramps have a duration of~$225 \,\text{ns}$, and are implemented by linearly sweeping the rf power and frequency to an acousto-optic modulator, following the procedure in Ref.~\cite{schine2022long}.
We note that the interaction times $t_\mathrm{int}$ quoted in this work do not include the duration of the ramps.

For each measurement in this work, we characterize the relevant parameters of the Rydberg drive $\hat{H}_\mathrm{Ryd}$: $\Omega_r$ and $\Delta$.
Since the Rydberg laser is locked to a high-finesse cavity, we control $\Delta$ directly by changing the rf frequency of a cavity offset lock.
The Rabi frequency~$\Omega_r$ is determined by driving on-resonance ($\Delta = 0$) Rabi oscillations for isolated single atoms.

For theoretical calculations (see \textit{Weak-dressing theory} and \textit{Strong-dressing theory}), we assume $C_6 \approx 2\pi\times 9.1\asciimathunit{GHz}  \, \upmu\text{m}^6$.
We estimate this value from experimental measurements of the two-photon $\ket{ee}$ to $\ket{rr}$ transition frequency for interatomic distances $r_{ij}$ between $3a_\mathrm{lat}$ and $7a_\mathrm{lat}$.
However, this measurement is susceptible to a variety of systematic effects, such as stray electric fields, which we do not characterize.
Therefore this value for $C_6$ may not be representative of Rydberg interactions in conditions that differ from those used in this work.

\subsection{Interaction potential~$V_{ij}$}
Figure~\ref{fig:1}b shows a measurement of the soft-core potential~$V_{ij}$ that describes the two-particle interactions in the system.
For this measurement, we initialize the atom array with a few isolated pairs of atoms at variable distance~$r_{01} = (1, 2, 3, 4, 5)a_\mathrm{lat}$.
We then apply the Rydberg-dressing pulse sequence (see II in Fig.~\ref{fig:1}c) followed by an additional~$\pi/2$ clock pulse.
A subsequent measurement of the atomic population corresponds to a measurement of the~$\hat{S}_z$ observable that oscillates with the frequency~$\omega$.
We extract~$\omega$ from a damped cosine fit and relate it to~$V(r_{01})$ by diagonalizing the two-particle Hamiltonian [$i,j=0,1$; see Eq.~\eqref{eq:ham}].
Finally, a numerical fit to $\omega  \left[ 1 + (r_{01}/\tilde{R}_b)^6\right]^{-1}$ yields the relevant fitted parameters $\tilde{V}_0=2\hbar \omega = h\times46.4(4)\kHz$ and $\tilde{R}_b=4.9(2)a_\mathrm{lat}$.
Notably, the interaction strength $V_0$  deviates significantly from the one obtained from the relations~$V_0 = \beta^3 \Omega_r \approx h\times80.6\kHz$ and the independently determined parameters $\Omega_r$ and $\Delta$.
This could be attributed to the relatively large $\beta = \Omega_r / (2\Delta) \approx 0.25$ employed throughout this work.
In particular, we find that the results from an exact-diagonalization calculation (see \textit{Strong-dressing theory}) are much closer to $\tilde{V}_0$.

\subsection{Wineland parameter $\xi_W^2$}
Each value of the Wineland parameter~$\xi_W^2$ shown in Fig.~\ref{fig:2}b and~d involves measuring the contrast~$C$ as well as the variance reduction $\sigma_\alpha^2 / \sigma_\mathrm{QPN}^2$.
Example measurements for these quantities are shown in Fig.~\ref{fig:2}a for the case of $N=4\times4=16$.
Note that the quantity~$\sigma_\mathrm{QPN}$ is calculated for the actual atom number obtained under the post-selection criterion explained in \textit{Post-selection}.
To reduce the statistical uncertainty of~$\xi_W^2$, we first obtain the optimal~$\alpha$ by a measurement and a numerical cosine fit like the one shown in Fig.~\ref{fig:2}a.
Subsequently, the value of the minimum variance reduction $\sigma_\alpha^2 / \sigma_\mathrm{QPN}^2$ is determined from an additional high-statistics measurement at the optimal $\alpha$.
We employ this procedure for all atom numbers except for $N=2\times2 = 4$, where the minimum variance reduction is obtained directly from a cosine fit.
For all atom numbers, we measure the Wineland parameter~$\xi_W^2$ for a range of different interaction times~$t_\mathrm{int}$.
To obtain the optimal~$\xi_W^2$ (and $t_\mathrm{int}$), we employ an empirical fit described by $a e^{-\Gamma_a t_\mathrm{int}} + b e^{-\Gamma_b t_\mathrm{int}}$ with fit parameters $a, b$ and $\Gamma_a, \Gamma_b$.
An example plot of this functional form together with experimental data can be found in Fig~\ref{fig:2}{b} (dark purple line).
The values of~$\xi_W^2$ plotted in Fig.~\ref{fig:2}d and quoted in the main text are obtained from the fit parameters and the resulting minimum of the above function.

\subsection{Atom-atom stability}
For the differential frequency comparisons shown in Fig.~\ref{fig:3}, we employ Ramsey spectroscopy.
At the end of the dark time~$T_{\text{dark}}$, we measure the signal
\begin{align}
    \small
    d^{(AB)}_z = \frac{C}{2}\left[\sin\left(\omega_{A} T_{\text{dark}}\right) - \sin\left(\omega_{B} T_{\text{dark}}\right)\right] \, 
\end{align}
where $\omega_{A} T_{\text{dark}}$ ($\omega_{B} T_{\text{dark}}$) is the phase accrued by ensemble $A$ ($B$) during the dark time. Since $(\omega_{A} - \omega_{B}) T_{\text{dark}}$ is generally small, $(\omega_{A} - \omega_{B}) \approx 2 d^{(AB)}_z / (C T_{\text{dark}})$.
We measure this angular frequency difference multiple times with a regular time interval of~$\approx 1.4\s$ between individual data points.
This allows us to obtain the differential stability between $A$ and $B$ by calculating the overlapping Allan deviation for a variable total averaging time~$\tau$ (see Fig.~\ref{fig:3}).
Note that we employ a low-gain digital servo to lock the laser onto the atomic resonance position during atom-atom stability measurements.
This servo ensures that the atomic populations remain near the optimal value by controlling the frequency of the clock laser beam using an acousto-optical modulator.

\subsection{Error bars and model fitting}
Throughout this work, fitted parameters corresponding to the CSS are extracted via maximum likelihood estimation under the assumption that the underlying distribution is binomial, whereas the corresponding parameters for the SSS are extracted via least-squares fits weighted by $1/N_i$, where $N_i$ is the number of atoms loaded on a given shot $i$ of the experiment.
For the contrasts contributing to Fig.~\ref{fig:2} and Fig.~\ref{fig:3}, confidence intervals are determined by non-parametric bootstrap using the basic method~\cite{carpenter2000bootstrap}.
Errors in other fitted parameters are determined from jackknifing, i.e., the displayed error bars correspond to the jackknife estimate for the standard error~\cite{efron1981jackknife}.
Unless noted otherwise, numerical least-squares fits weight the data points with their inverse variance.
For the data corresponding to the variance reduction in Fig.~\ref{fig:2}a, error bars of the variance are also determined by jackknifing.

\subsection{Weak-dressing theory}
For the weak-dressing theory shown in Fig.~\ref{fig:2}d, we directly employ the analytical results given in Ref.~\cite{gil2014spin} to obtain the relevant experimental parameters $C$, $\sigma_\alpha^2/\sigma_\mathrm{QPN}$, and $\xi_W^2$.
We note that the ramps of the Rabi frequency and detunings (see \textit{Rydberg drive and parameters}) are neglected in these calculations.
The atom numbers for the weak-dressing theory curve shown in Fig.~\ref{fig:2}{d} (light purple line) are $N=1\times 2$, $1\times3$, $2\times2$, $\ldots$, $5\times 5$, and $5\times m$ with $m=6,7,\ldots,20$.
Here, the spacing between atoms is set to $2a_\mathrm{lat}$.
While the independently determined experimental parameters $\Omega_r$ and $\Delta$ in each of the measurements aggregated in Fig.~\ref{fig:2}{d} slightly differ, the weak-dressing theory is calculated for the parameters of the $N=4 \times 4 = 16$ data set.

Following Ref.~\cite{worm2013relaxation}, we find the following expression for the relevant two-particle correlator shown in Fig.~\ref{fig:2}c (spatial correlations),
\begin{align}
    \small
    g^{(2)}_{ij} 
    &= \frac{1}{4}\left(P_{ij}^- - P_{ij}^+\right) \sin^2\alpha \nonumber
     + \frac{1}{4}\sin\alpha\cos\alpha\sin\varphi_{ij}\\
     &\quad\times \left( \prod_{k\neq i, j} \cos\varphi_{ik} + \prod_{k\neq i, j} \cos\varphi_{jk}\right)
\end{align}
with the expression $P_{ij}^\pm = \frac{1}{2}\prod_{k\neq i, j}\cos(\varphi_{ki} \pm \varphi_{kj})$ and the interaction phase $\varphi_{ij} = V_{ij} t_\mathrm{int} / (2\hbar)$.
The above quantity is then related to $g_\mathbf{r}^{(2)}$ mentioned in the main text using the distances $r_{ij}$ between atoms in the experimentally prepared subarray with atom number $N=5\times14=70$.
Here, the spacing between atoms is set to $(3, 2)a_\mathrm{lat}$ along the $x$ and $y$ axis to match the experimental realization.
The theory shown in Fig.~\ref{fig:2}c is calculated at the optimal interaction time by first minimizing $\xi_W^2$ calculated with the weak-dressing theory from Ref.~\cite{gil2014spin}.

\subsection{Strong-dressing theory}
For small atom numbers ($N\leq9$ in this work), the full three-level system ($\ket{g}$, $\ket{e}$, and $\ket{r}$) is simulated via exact diagonalization.
In this case, the Hamiltonian describing the off-resonantly driven Rydberg system is:
\begin{align}
    \small
    &\frac{\hat{H}_3}{\hbar} = \frac{\Omega_r(t)}{2}  \sum_i \left( \ket{e}\bra{r}_i + \text{h.c.} \right) + \Delta(t) \sum_i \ket{r}\bra{r}_i \\
    &\quad + \sum_{i<j} \frac{C_6}{r_{ij}^6} \ket{r}\bra{r}_i \ket{r}\bra{r}_j
    + \frac{\Omega_c(t)}{2}  \sum_i \left( \ket{g}\bra{e}_i + \text{h.c.} \right). \nonumber
\end{align}
We implement a step function for the clock Rabi frequency~$\Omega_c(t)$ and a linear ramp both for the detuning~$\Delta(t)$ and Rydberg Rabi frequency~$\Omega_r(t)$ to model the experimental procedures (see \textit{Rydberg drive and parameters}).
These linear ramps have a duration of $225\,\mathrm{ns}$, and are discretized with a step size of $6.5\,\mathrm{ns}$ in our simulations.
The initial state is given by $\ket{\psi_0} = \ket{gg\ldots g}$ and we time-propagate it under $e^{-i\hat{H}_3 t / \hbar}$ using the software library \texttt{quspin}~\cite{weinberg2017quspin}.
From the final state, the relevant quantities $C$, $\sigma_\alpha^2 / \sigma_\mathrm{QPN}^2$, and $\xi_W^2$ are determined and compared to the experimental data~\cite{seeSupplement}.
Note that this calculation assumes perfect clock rotations and does not contain any free parameters, i.e., the relevant parameters of $\hat{H}_3$ are determined independently.

\subsection{Ellipse fitting}
For MLE in Fig.~\ref{fig:4}, we must model noise about the mean excitation fractions of the two ensembles.
As mentioned in the main text, one challenge associated with ellipse fitting is that we lack a detailed understanding of the noise distribution for the SSS.
In this section, we introduce an empirical model for SSSs which we use to perform MLE.
However, we note that future theoretical work will be required in order to assess the accuracy of this model.

The empirical model we use to fit data in the main text is defined by
\begin{align}
    \begin{split}
    \small
    &f\left(p_A, p_B | \phi, C, y_0, \vec{\zeta} \, \right) =\\& \frac{1}{2\pi} \int_0^{2\pi} f_{\theta} \left(p_A, p_B | \phi, C, y_0, \vec{\zeta} \, \right) \, \text{d}\theta \, .
    \end{split}
\end{align}
Here, $p_{A,B}$ are specific measurement values for the observables $\hat{P}_{A,B} = \mathds{1}/2 + \hat{S}_z^{(A,B)} / N_{A,B}$, which correspond to the mean excitation fraction in each ensemble. Additionally, $f_{\theta}$ is the probability mass function for the two ensembles with a specified atom-laser phase $\theta$, and takes the form
\begin{align}
\begin{split}
    \small
    &f_{\theta}\left(p_A, p_B | \phi, C, y_0, \vec{\zeta} \, \right) = \\
    \frac{1}{\mathcal{N}_A \mathcal{N}_B} &\Biggl{[}\begin{pmatrix} N \\ k_A \end{pmatrix}  P_A^{k_A} (1 - P_A)^{N - k_A}\Biggr{]}^{1/\zeta^2(\theta)} \\
     \times &\Biggl{[} \begin{pmatrix} N \\ k_B \end{pmatrix}  P_B^{k_B} (1 - P_B)^{N - k_B} \Biggr{]}^{1/\zeta^2(\theta + \phi)} \, .
\end{split}
\end{align}
In this equation, we take $N_A, N_B = N$, $k_{A,B} = p_{A,B} N$, and $P_{A,B} = \left\langle \hat{P}_{A,B} \right\rangle$ with
\begin{align}
\begin{split}
    \small
    P_A &= \frac{C}{2} \cos(\theta) + y_0 \\
    P_B &= \frac{C}{2} \cos(\theta + \phi) + y_0 \, .
\end{split}
\end{align}
We refer to $\theta$ as the atom-laser phase and $\phi$ as the differential phase; $y_0$ is an offset. We take $C$ and $y_0$ to be the same for both ensembles. Finally, $\mathcal{N}_{A,B}$ are normalization factors, and
\begin{align}
    \small
    \zeta^2 (\theta) = \zeta_{0}^2 \sin^2\left(\theta \right) + \zeta_{1}^2 \cos^2\left(\theta \right) \,.
\end{align}
We note that when $\vec{\zeta} = (1, 1)$, $f_{\theta}$ is simply the product of binomial distributions and representative of a CSS.
By design, $\zeta_0$ then plays an analogous role to the squeezing parameter in the large-$N$ limit, where -- by the central limit theorem for the binomial distribution -- $f_{\theta}$ converges to the product of normal distributions [as long as $P_A, P_B \in (0,1)]$.
However, we emphasize that the model parameters $\vec{\zeta} = (\zeta_0, \zeta_1)$ do not directly correspond to the squeezing or anti-squeezing present in the state.
Nevertheless, we define the likelihood function
\begin{align}
    \small
    \mathcal{L}\left( \phi, C, y_0, \vec{\zeta} | A \right) = \prod_{i\in I} f\left(p^{(i)}_A, p^{(i)}_B | \phi, C, y_0, \vec{\zeta} \, \right)
\end{align}
where $p^{(i)}_A, p^{(i)}_B$ are the measurement results for a given trial index $i$ in the full set of measurement indices $I$, and $A = \{ (p_A^{(i)}, p_B^{(i)}) | i \in I \}$ is the corresponding set of measurements.

To extract the precision with which we are able to infer the parameter $\phi$, we split our data into a calibration data set $A_{\text{cal}} = \{ (p_A^{(i)}, p_B^{(i)}) | i \in I_{\text{cal}} \}$, and a measurement data set $A_{m} = \{ (p_A^{(i)}, p_B^{(i)}) | i \in I_{m} \}$.
Here, the set $I_{\text{cal}}$ contains a random selection of half of the indices $i$ from 0 to $n-1$, where $n$ is the total number of measurements, and $I_{m}$ contains the remaining indices.
For consistency, we use the same random samplings for both the SSS and CSS.
The role of the calibration data set is to extract estimates of the parameters $C$, $y_0$, and $\vec{\zeta}$, as well as the uncertainty in these estimates.
We characterize these uncertainties using non-parametric bootstrap, and resample the data in $A_{\text{cal}}$ a total of 50 times.
For a given bootstrap sample $A_\text{B}$, maximizing the likelihood $\mathcal{L}( \phi, C, y_0, \vec{\zeta} | A_\text{B} )$ yields a set of inferred parameters $(\phi_{\text{cal}}, C_{\text{cal}}, y_{\text{cal}}, \vec{\zeta}_{\text{cal}})$.
We discard $\phi_{\text{cal}}$, and construct a new likelihood function for each $(C_{\text{cal}}, y_{\text{cal}}, \vec{\zeta}_{\text{cal}})$:
\begin{align}
    \small
    \mathcal{L}^{\text{B}}\left( \phi|A \right) = \mathcal{L}\left( \phi, C_\text{cal}, y_{\text{cal}}, \vec{\zeta}_\text{cal} | A \right).   
\end{align}

We use the corresponding likelihood function calibrated by the original (not resampled) data set $A_{\text{cal}}$ to extract the value of $\phi$, along with its statistical variance and Allan deviation, from the measurement data $A_m$. 
Repeating this procedure using each $\mathcal{L}^{\text{B}}\left( \phi|A_m \right)$ allows us to estimate the effect of calibration errors in the secondary parameters $C, y_0,$ and $\vec{\zeta}$, independent of statistical uncertainty in the measurement data.
The error bars appearing in Fig.~\ref{fig:4}\subfigref{c} correspond to the quadrature sum of these calibration errors with the statistical uncertainty.

To compute the overlapping Allan deviation for each bootstrap sample of calibration parameters, we follow the recipe for jackknifing described in \cite{marti2018imaging} and calculate the amount that each measurement $i$ pulls the overall phase estimate:
\begin{align}
    \small
    \phi^{\text{JK}}_i = \frac{n}{2}\phi' - \left(\frac{n}{2} - 1 \right)\phi_{\neq i}' \, 
\end{align}
where $n/2$ is the number of points in the measurement set $A_m$, and $\phi'$ and $\phi_{\neq i}'$ are defined as:
\begin{align}
\begin{split}
    \small
    \phi_{\neq i}' &= \text{argmax}_{\phi \in [0,\pi]} \, \left[ \mathcal{L}^{\text{B}} (\phi | A_{m}\setminus i)\right] \\
    \phi' &= \text{argmax}_{\phi \in [0,\pi]} \, \left[ \mathcal{L}^{\text{B}} (\phi | A_m)\right] .
\end{split}
\end{align}
Here, $A_{m}\setminus i$ refers to the data set $A_{m}$ with element $(p_A^{(i)}, p_B^{(i)})$ removed.

The stabilities in Fig.~\ref{fig:4}\subfigref{c} are calculated by taking the overlapping Allan deviation of the single-shot estimates of the differential phase, $\phi^{\text{JK}}_i$.
We plot the Allan deviation as a function of the number of measurements $m$ (see Fig.~\ref{fig:4}\subfigref{c}), which is related to the total measurement time~$\tau$ by $\tau = m\, T_{\text{cycle}}$, where $T_{\text{cycle}}$ is the cycle time of the experiment.
For a direct comparison between the SSS and CSS, we use the same procedure and model to compute the Allan deviation in each case, and plot the ratio in Fig.~\ref{fig:4}\subfigref{c}.
Since the distribution $f$ converges to the model for a CSS in the limit where $\zeta_0 = \zeta_1 = 1$, this procedure should not underfit CSS data.

\bibliography{references}

\end{document}


\title{Supplementary information: Realizing spin squeezing with Rydberg interactions in a programmable optical clock}

\author{William J. Eckner, Nelson \surname{Darkwah Oppong}, Alec Cao, Aaron W. Young, William R. Milner, John M. Robinson, Jun Ye, Adam M. Kaufman}

\affiliation{%
 JILA, University of Colorado and National Institute of Standards and Technology,
and Department of Physics, University of Colorado, Boulder, Colorado 80309, USA
}%

\date{\today}

\maketitle
\tableofcontents
\newpage

\section{Characterization of local $\hat{\sigma}_z$ operations}

\begin{figure}
    \includegraphics[width=0.67\textwidth]{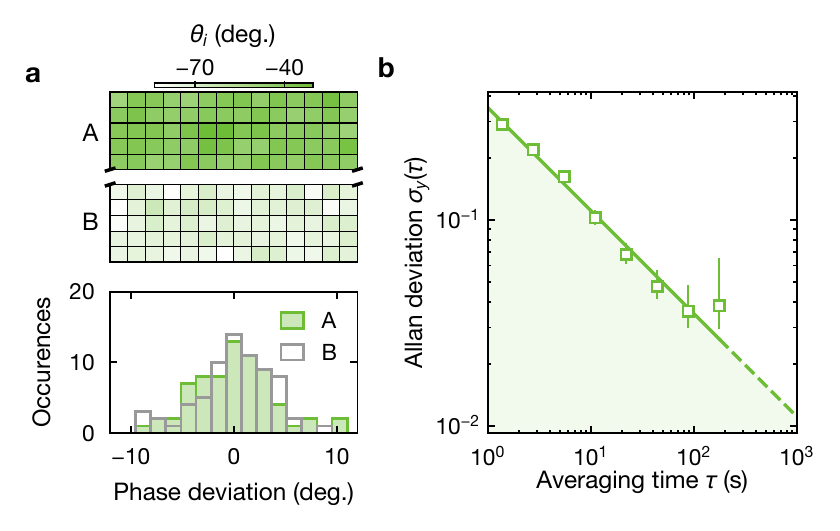}
    \caption{\label{fig:phaseshift} \textbf{Characterization of tweezer phase shifting homogeneity and stability.} \subfiglabel{a}~Site-resolved Ramsey fringe phases for two $N=5 \times 14$ subarrays with the phase shifting protocol described in Fig.~4\subfigref{a} applied during the dark time to ensemble $A$.
    For each site $i$, $\theta_i$ is obtained by fitting $P_i \sim \sin (\theta_L + \theta_i)$.
    The lower panel shows a histogram of the difference of $\theta_i$ from the relevant average subarray phase $\theta_{A}$ and $\theta_{B}$.
    The standard deviation of the $\theta_i$ distribution for ensemble $A$ ($B$) is $3.8^\circ$ ($3.6^\circ$).
    \subfiglabel{b}~Overlapping Allan deviation for the applied phase shift.
    The laser phase $\theta_L$ is set so that $(P_A + P_B)/2 \approx 1/2$ on average.
    The deviation is computed for $2 d_z^{(AB)}/(\phi C)$; here, $\phi$ and $C$ are computed from the measurement in \textbf{a}, with $\phi = \theta_A - \theta_B = 33.2(1.3)^{\circ}$ and $C = 0.971(2)$ obtained from averaging the contrasts of subarrays $A$ and $B$.}
\end{figure}

In order to minimize added technical noise, the local light-shifting protocol presented in Fig.~4a and the Methods should be the same across the atom array (spatially homogeneous), and consistent between different experiments (temporally homogeneous).
To characterize spatial homogeneity, we measure site-resolved Ramsey fringes, and analyze the distribution of fitted phases, shown in Fig.~\ref{fig:phaseshift}a.
We benchmark the temporal homogeneity of the the protocol by performing an atom-atom stability measurement for $5 \times 14$ subarrays with the tweezers applied during a short dark time.
Because the ensembles are phase shifted relative to each other, the phase of the final $\pi/2$-pulse is chosen such that the average $S_z^{(A)} + S_z^{(B)} = 0$. Results of this measurement are shown in \ref{fig:phaseshift}b. 
We note that any ellipse measurements utilizing $\hat{\sigma}_z$ operations alternate experimental shots where the Rydberg laser is applied or not, and thereby interleave data acquisition for the SSS and CSS.
As a result, the SSS and CSS should experience similar systematic effects, such as any slow residual drift of the applied phase offset between subarrays. 

\section{Signatures of collective dissipation}

\begin{figure}
\includegraphics[width=0.67\linewidth]{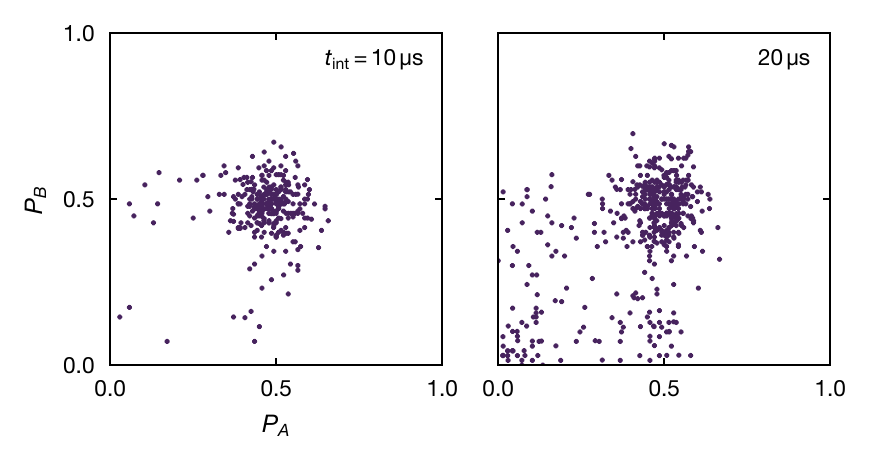}
\caption{
\label{fig:coll_loss}
\textbf{Emerging collective effects at late times.}
As in the main text, we denote the excitation fraction $P_A, P_B$ for subarray $A$, $B$, respectively. In the left (right) panel we parametrically plot the excitation fractions $P_A$ and $P_B$, measured after a total interaction time $t_{\text{int}} = 10 \, \mu \rm{s}$ ($20 \, \mu \rm{s}$).
For this measurement, we determine the excitation fraction immediately after the second Rydberg-interaction pulse (see Fig.~1c in the main text), and with two $N=5\times 14$ subarrays. 
}
\end{figure}

When increasing the interaction time to $t_\mathrm{int}\geq 10\us$ -- a regime beyond the typical optima for preparing SSSs -- signs of a bimodal distribution emerge in the atomic excitation fraction for two $N=5\times 14$ subarrays.
By plotting the excitation fractions $P_A, P_B$ for two subarrays $A$ and $B$, respectively, as shown in Fig.~\ref{fig:coll_loss}, we can further see a bimodal distribution emerge in $P_A$, even when controlling for $P_B$ (and vice-versa).
This observation indicates the presence of collective effects and, therefore, appears to be consistent with the presence of collective dissipation phenomena, which have been studied in similar Rydberg-atom-array platforms~\cite{zeiher2016many, boulier2017spontaneous, young2018dissipation, guardado2021quench, festa2022blackbody}.

\section{Derivation of QPN and $\xi_W^2$}

In this section, we derive expressions for quantum projection noise (QPN) and the Wineland parameter presented in the main text. To begin, $S_z / N$ for two ensembles~$A$ and~$B$ in a Ramsey-style measurement will be given by
\begin{align}
\begin{split}
    \frac{S_z^{(A)}}{N} &= \frac{C}{2} \sin(\theta + \pi/2) + y_A \\
    \frac{S_z^{(B)}}{N} &= \frac{C}{2} \sin(\theta + \phi + \pi/2) + y_B
\end{split}
\end{align}
where we refer to $\theta$ as the atom-laser phase, and $\phi$ as the differential phase. We have furthermore assumed that both ensembles have the same contrast $0 \le C \le 1$, but potentially different offsets $C-1 \le 2\,  y_A, 2\, y_B \le 1 - C$. Finally, $N$ refers to the atom number in each ensemble, which we take to be equal ($N_A = N_B = N$).

We are interested in the noise in the measurement of $\hat{d}_z^{(AB)}$, as defined in the main text. To motivate this decision, we show that this provides a measurement of the differential phase~$\phi$. Writing out the expectation value of $\hat{d}_z^{(AB)}$ gives
\begin{align}
\begin{split}
    d_z^{(AB)} = \frac{S_z^{(A)} - S_z^{(B)}}{N} = \frac{C}{2} [\sin(\theta + \pi/2) - \sin(\theta + \phi + \pi/2)] + (y_A - y_B).
\end{split}
\end{align}
Taylor-expanding about the point $(\theta, \phi) = (\pi/2,0)$ yields
\begin{align}
\begin{split}
    d_z^{(AB)} \approx \frac{C}{2} \phi + (y_A - y_B).
\end{split}
\end{align}
The phase uncertainty of our measurement will then be
\begin{align}
\begin{split}
    \Delta \phi &= \Delta \hat{d}_z^{(AB)} {\left| \frac{\text{d} d_z^{(AB)} }{\text{d} \phi} \right|}^{-1}
    = \frac{2 \Delta \hat{d}_z^{(AB)}}{C}.
\end{split}
\end{align}
We now define the quantum-projection-noise limit as the variance of an ideal (contrast $C=1$) coherent spin state $\ket{\text{CSS}}$
\begin{align}
    \ket{\text{CSS}} &= \ket{\theta + \phi}_A \otimes \ket{\theta}_B
\end{align}
where
\begin{align}
\begin{split}
    \ket{\theta}_A &= \bigotimes_{i = 0}^{N-1} [e^{-i\theta/2} \ket{e}_i + e^{i \theta/2} \ket{g}_i] / \sqrt{2} \\
     \ket{\theta + \phi}_B &= \bigotimes_{j = 0}^{N-1} [e^{-i(\theta + \phi)/2} \ket{e}_j + e^{i (\theta + \phi)/2} \ket{g}_j] / \sqrt{2} \, .
\end{split}
\end{align}
and $i$ ($j$) indexes atoms in ensemble $A$ ($B$). Using the notation presented in the main text,
\begin{align}
    \sigma_{\text{QPN}}^2 &= \bra{\text{CSS}} [\hat{d}_z^{(AB)}]^2\ket{\text{CSS}} - [\bra{\text{CSS}} \hat{d}_z^{(AB)} \ket{\text{CSS}}]^2 \\
    &= \frac{1}{2N} \, .
\end{align}
From this, the standard quantum limit on our estimation of $\phi$ is 
\begin{align}
    (\Delta \phi)_{\text{SQL}} = \sqrt{\frac{2}{N}} \, .
\end{align}
Next, we derive an expression for the squeezing parameter, which is typically defined in the context of a Ramsey measurement with a single ensemble of $N$ atoms and an atom-laser phase $\theta$ as \cite{wineland1992spin}
\begin{align}
    \xi_{\theta}^2 = \frac{(\Delta \theta)^2}{(\Delta \theta)^2_{\text{SQL}}} \, .
\end{align}
Here, $\Delta \theta$ is the uncertainty in the measured atom-laser phase, and 
\begin{align}
    (\Delta \theta)_{\text{SQL}} = \sqrt{\frac{1}{N}} \, .
\end{align}
In the context of our differential measurement of $\phi$, we therefore define the parameter $\xi^2_{\phi}$ as 
\begin{align}
\begin{split}
    \xi^2_{\phi} = \frac{(\Delta \phi)^2}{(\Delta \phi)^2_{\text{SQL}}} &=\frac{\sigma_{\alpha}^2}{C^2} \frac{1}{\sigma^2_{\text{QPN}}}
\end{split}
\end{align}
where the angle $\alpha$ sets the measurement quadrature. The Wineland parameter $\xi_W^2$ in the main text is then
\begin{align}
    \xi_W^2 = \left[ \xi^2_{\phi} \right]_{\text{min}} &=\frac{\sigma_{\text{min}}^2}{C^2} \frac{1}{\sigma^2_{\text{QPN}}}
\end{align}
where `min' refers to the minimum over $\alpha$.

\section{Fisher information of CSS for ellipse fitting}

\begin{figure}
    \centering
    \includegraphics[width=0.67\textwidth]{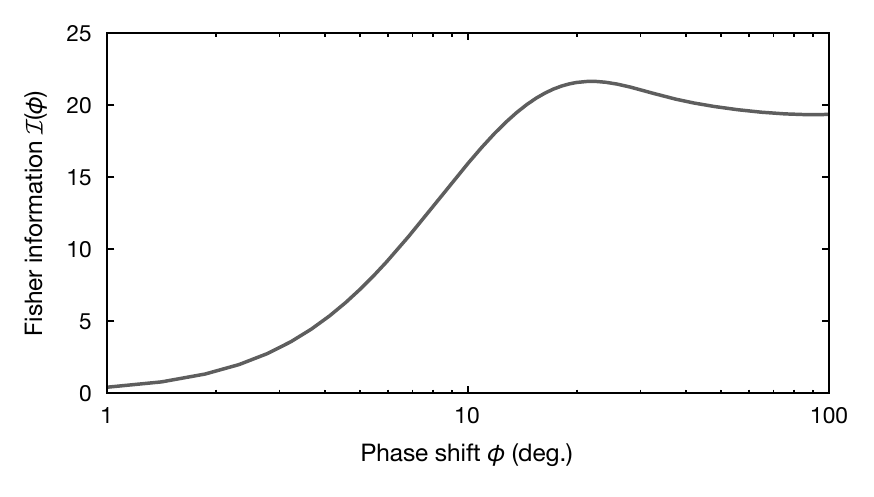}
    \caption{\textbf{Measurement sensitivity of ellipse fitting.}
    Here, we consider a variable phase shift $\phi$ between two subarrays and quantify the measurement sensitivity using the classical Fisher information (solid gray line).
    Here, a coherent spin state with contrast $C=0.95$ is considered.
    }
    \label{fig:fisherinfo}
\end{figure}

In Fig.~\ref{fig:fisherinfo}, we present a calculation of the Fisher information in a coherent spin state (CSS) for measuring $\phi$ in an ellipse-fitting protocol. In this calculation, we consider the probability mass function 

\begin{align}
    f_{\text{CSS}}\left(p_A, p_B | \phi, C, y_0 \right) = f\left(p_A, p_B | \phi, C, y_0, \vec{\zeta} = (1, 1) \, \right)
\end{align}
where the function $f\left(p_A, p_B | \phi, C, y_0, \vec{\zeta} \, \right)$ is defined in the Methods. The parameter regime in which $\vec{\zeta} = (1, 1)$ corresponds to a binomial model, which captures the statistics of uncorrelated atoms in a CSS. The Fisher information for a measurement of $\phi$ with this probability distribution is then
\begin{align}
    \mathcal{I}_{\text{CSS}}\left( \phi_0 \right) = \sum_{p_A = 0/N}^{N/N} \sum_{p_A = 0/N}^{N/N} \left\{ \frac{\partial}{\partial \phi} \log \left[  f_{\text{CSS}}\left(p_A, p_B | \phi, C, y_0 \right) \right] \Big{|}_{\phi = \phi_0} \right\}^2  f_{\text{CSS}}\left(p_A, p_B | \phi_0, C, y_0 \right) \, .
\end{align}
In Fig.~\ref{fig:fisherinfo}, we plot $\mathcal{I}_{\text{CSS}}\left( \phi \right)$ versus $\phi$ for the parameters $C = 0.95$, $y_0 = 0.5$, which are representative of typical experimental values.

\begin{figure}
    \centering
    \includegraphics[width=\textwidth]{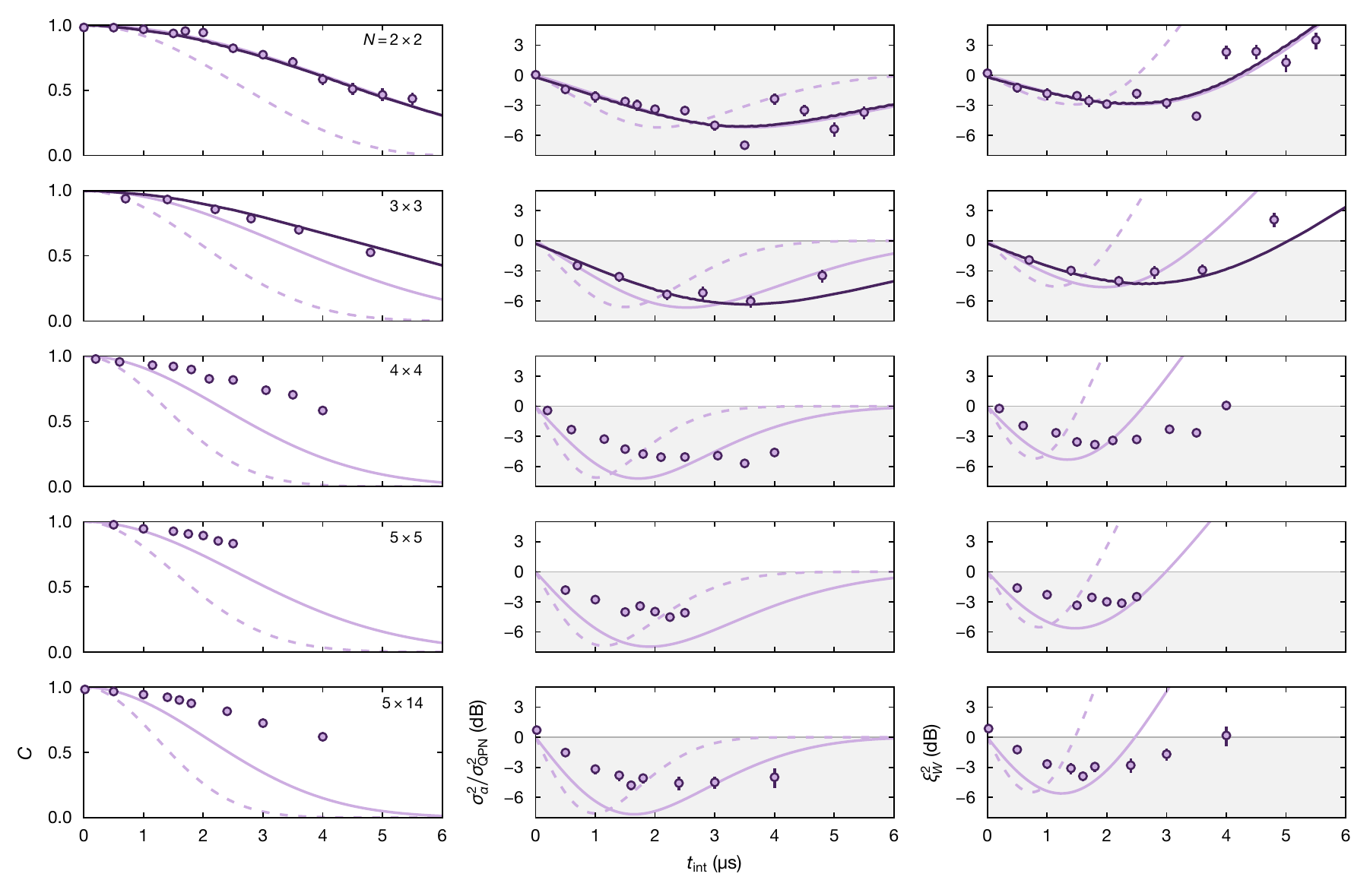}
    \caption{\textbf{Squeezing dynamics in the experiment and numerical simulations.}
    Here, we show experimental measurements (purple circles) for the contrast $C$ (left column), variance reduction $\sigma_\alpha^2 / \sigma_\mathrm{QPN}^2$ (center column), and the Wineland parameter $\xi_W^2$ (right column).
    From top to bottom, the atom number increases from $N=2\times2 = 4$ to $5\times14 = 70$ in each row.
    Dark purple lines correspond to an exact-diagonalization calculation (see Methods) for the parameters in the experiment.
    Solid light purple lines show the theoretical predictions from weak-dressing~\cite{gil2014spin} using $\tilde{V}_0$ and $\tilde{R}_b$ from the fit shown in Fig. 1b of the main text.
    Dashed purple lines correspond to the same theory, but with $V_0 =\hbar \beta^3 \Omega_r$ and $R_b = {|C_6 / (2\Delta)|}^{(1/6)}$.
    }
    \label{fig:overview}
\end{figure}

\section{Comparison of dynamics in the experiment with theory}

In this section, we directly compare the dynamics observed in the experiment to theoretical predictions from numerical simulations.
To this end, we consider the weak and strong-dressing theory (exact diagonalization), as outlined in the Methods section of the main text.
While the exact diagonalization directly employs the independently determined parameters $\Omega$, $\Delta$, and $C_6$, we choose two different approaches for the weak-dressing theory.
In the first approach, we use the values of $\tilde{V}_0$ and $\tilde{R}_b$ obtained from the fit shown in Fig.~1b with a minor rescaling to adjust for slightly different parameters in each measurement.
In the second approach, we calculate $V_0 = \hbar\beta^3 \Omega_r$ and $R_b = {|C_6 / (2\Delta)|}^{(1/6)}$ from $\Omega_r$, $\Delta$, and $C_6$.

For the exact-diagonalization calculation (see dark purple lines in Fig.~\ref{fig:overview}), we generally find good qualitative agreement between experiment and theoretical prediction for the numerically accessible subarray sizes $N=2\times2$ and $3\times 3$.
For the first weak-dressing approach (see light purple lines in Fig.~\ref{fig:overview}), we find similarly good agreement for $N=2\times 2$, but for larger subarray sizes the dynamical time scale becomes significantly faster than the one observed in the experiment.
Moreover, the theoretically predicted maximum squeezing~$1/\xi_W^2$ becomes significantly larger than the experimentally observed one.
For the second weak-dressing approach (see dashed light purple lines in Fig.~\ref{fig:overview}), the theoretically predicted dynamics are even faster since $\tilde{V_0} < V_0$ which sets the characteristic time scale of the system~\cite{gil2014spin}.

\bibliography{references}